# Stochastic Flexibility Evaluation for Virtual Power Plant by Aggregating Distributed Energy Resources

Siyuan Wang, *Student Member, IEEE*, Wenchuan Wu, *Senior Member, IEEE*

*Abstract*—To manage huge amount of flexible distributed energy resources (DERs) in the distribution networks, the virtual power plant (VPP) is introduced in industry. The VPP can optimally dispatch these resources in a cluster way and provide flexibility for the power system operation as a whole. Most existing works formulate the equivalent power flexibility of the aggregating DERs as deterministic optimization models without considering their uncertainties. In this paper, we introduce the stochastic power flexibility range (PFR) and time-coupling flexibility (TCF) to describe the power flexibility of VPP. In this model, both operational constraints and the randomness of DERs' output are incorporated, and a combined model and data-driven solution is proposed to obtain the stochastic PFR, TCF and cost function of VPP. Finally, a numerical test was performed. The results show that the proposed model not only has higher computational efficiency than the scenario-based methods, but also achieve more economic benefits.

## NOMENCLATURE

### A. Parameters

| | |
|---|---|
| $\boldsymbol{K}, \boldsymbol{G}, \boldsymbol{J}, \boldsymbol{b}, \boldsymbol{c}, \boldsymbol{d}_{ij}$ | Network parameters matrices and vectors used to calculate the multi-phase unbalanced power flow |
| $\boldsymbol{A}_{i,s}^{\psi}, \boldsymbol{b}_{i,s}^{\psi}$ | Parameters of polygonal power control capability charts of DER $s$ at phase $\psi$ of bus $i$, where $s$ can denote combined heat and power generator (CHP), photovoltaic generator (PV), energy storage system (ESS), or wind turbine (WT) |
| $\bar{P}_{i,\mathrm{PV}}^{\psi,t,\max}, \bar{P}_{i,\mathrm{WT}}^{\psi,t,\max}$ | Forecast value of output power of photovoltaic generator and wind turbine at phase $\psi$ of bus $i$ at time $t$ |
| $\bar{P}_{i,\mathrm{load}}^{\psi,t}$ | Forecast value of active load power at phase $\psi$ of bus $i$ at time $t$ |
| $\varphi_{i,\mathrm{load}}^{\psi}$ | Power factor of load at phase $\psi$ of bus $i$ at time $t$ |
| $V_i^{\psi,\max}, V_i^{\psi,\min}$ | Maximum and minimum voltage amplitude at phase $\psi$ of bus $i$ |
| $I_{ij}^{\psi,\max}$ | Maximum current amplitude at phase $\psi$ of branch $ij$ |
| $\alpha^{V+}, \alpha^{V-}$ | Risk probability of voltage amplitude exceeding the upper and lower limits |
| $\alpha^{I+}, \alpha^{I-}$ | Risk probability of positive and negative currents amplitude exceeding the limits of branches |
| $\alpha^P$ | Risk probability of maximum total output active power lower than the evaluated upper bounds |
| $r_{i,\mathrm{CHP}}$ | Ramping rate of CHP at bus $i$ |
| $\alpha_{i,\mathrm{ESS}}, E_{i,\mathrm{ESS}}^{\min}, E_{i,\mathrm{ESS}}^{\max}$ | Self-discharging rate, minimum and maximum energy constraints of ESS at bus $i$ |
| $a_{i,\mathrm{CHP}}, b_{i,\mathrm{CHP}}, c_{i,\mathrm{CHP}}$ | Quadratic cost coefficients of CHP at bus $i$ |
| $K_i^{\mathrm{dis}}, K_i^{\mathrm{ch}}$ | Discharge and charge cost parameter of ESS of bus $i$ |
| $\Delta t$ | Time interval of one period |
| $P_{\mathrm{PCC}}^{t,(k)}, C_{\mathrm{VPP}}^{t,(k)}$ | The $k$-th sample of active output power of VPP and corresponding minimum cost |

### B. Variables

| | |
|---|---|
| $\boldsymbol{s}_i^Y, \boldsymbol{s}_i^\Delta$ | The complex injection power of wye- and delta-connection sources at bus $i$ |
| $\boldsymbol{s}^Y, \boldsymbol{s}^\Delta$ | All the wye- and delta-connection sources in the VPP |
| $\boldsymbol{V}$ | Voltage amplitudes vector of all phases of all the buses |
| $s_0$ | Complex power injection at PCC |
| $\boldsymbol{i}_{ij}$ | Branch current vector of all phases of all branch |
| $P_{i,s}^{\psi,t}, Q_{i,s}^{\psi,t}$ | Active and reactive output power of DER $s$ at phase $\psi$ of bus $i$ at time $t$ |
| $E_{i,\mathrm{ESS}}^t$ | Energy storage of ESS at phase $\psi$ of bus $i$ at time $t$ |
| $\tilde{P}_{i,\mathrm{PV}}^{\psi,t,\max}, \hat{P}_{i,\mathrm{PV}}^{\psi,t}$ | Real-time maximum output power and curtailed power of photovoltaic generator at phase $\psi$ of bus $i$ at time $t$ |
| $\tilde{P}_{i,\mathrm{WT}}^{\psi,t,\max}, \hat{P}_{i,\mathrm{WT}}^{\psi,t}$ | Real-time maximum output power and curtailed power of wind turbine at phase $\psi$ of bus $i$ at time $t$ |
| $\tilde{e}_{i,\mathrm{PV}}^{\psi,t}, \tilde{e}_{i,\mathrm{WT}}^{\psi,t}$ | Forecast error of PV or WT at phase $\psi$ of bus $i$ at time $t$ |
| $\tilde{P}_{i,\mathrm{load}}^{\psi,t}, \tilde{Q}_{i,\mathrm{load}}^{\psi,t}$ | Real-time active and reactive power of load at phase $\psi$ of bus $i$ at time $t$ |
| $\tilde{P}_{i,\mathrm{inj}}^{\psi,t}, \tilde{Q}_{i,\mathrm{inj}}^{\psi,t}$ | Active and reactive injection power of load at phase $\psi$ of bus $i$ at time $t$ |
| $\boldsymbol{p}^{Y,t}, \boldsymbol{q}^{Y,t}$ | Active and reactive injection power vectors of all wye-connection phases at time $t$ |
| $\boldsymbol{p}^{\Delta,t}, \boldsymbol{q}^{\Delta,t}$ | Active and reactive injection power vectors |

---
†This work was supported in part by the National Science Foundation of China under Grant 51725703. *(Corresponding author: Wenchuan Wu.)* e-mail: wuwench@tsinghua.edu.cn)
S. Wang and W. Wu are with the State Key Laboratory of Power Systems, Department of Electrical Engineering, Tsinghua University, Beijing 100084, China.

| | |
|---|---|
| | of all delta-connection phases at time $t$ |
| $\tilde{V}_i^{\psi,t}$ | Voltage amplitudes at phase $\psi$ of bus $i$ at time $t$ |
| $\tilde{I}_{ij}^{\psi,t}$ | Branch current at phase $\psi$ of branch $ij$ at time $t$ |
| $x^t$ | Vector collected all the decision variables at time $t$ |
| $P_{PCC}^t, Q_{PCC}^t$ | Active and reactive injection power at PCC of time $t$ |

*C. Notation and functions*

| | |
|---|---|
| $j$ | The imaginary unit |
| $\mathcal{T}$ | Sets of all the time slots |
| $\mathcal{N}_Y, \mathcal{N}_\Delta$ | Sets of wye- and delta-connection buses |
| $Y, \Delta$ | Sets of wye- and delta-connection phases |
| $\mathcal{N}(x;\mu,\Sigma)$ | The probability density function of multivariable Gaussian distribution with $\mu$ as the expectation vector; and $\Sigma$ as the covariance matrix |
| $PDF_{\tilde{x}}(x)$ | The probability density function (PDF) of multivariable $\tilde{x}$ |
| $CDF_{\tilde{x}}(x)$ | The cumulative density function (CDF) of variable $\tilde{x}$ |
| $\Phi(x)$ | CDF of standard Gaussian distribution |
| $Quant(\alpha \mid \tilde{x})$ | Quantile of variable $\tilde{x}$ at probability $\alpha$ |
| $ReLU(\cdot)$ | Rectified Linear Unit function |

I. INTRODUCTION

*A. Motivation*

High penetration of renewable energy in distribution networks bring new challenges in terms of voltage violation, power quality deterioration, protection relay failure and insufficient flexibility. On the one hand, it is very difficult to control thousands or even millions of small DERs directly. On the other hand, these resources cannot participate in the electricity market individually due to their small capacity. The concept of virtual power plant (VPP) provides a promising solution to this problem [1].

VPP is a collection of distributed generators (DGs), distributed energy storage units and controllable loads. It uses advanced regulation and communication technologies to manage these DERs in a cluster way [2], [3]. There are two types of VPP, one is technical VPP (TVPP) and the other is commercial VPP (CVPP). TVPP is responsible for secure and optimal operation of VPP. All the DERs in a TVPP can be considered as a whole to dispatch and control. Considering the technique constraints, TVPP dispatches the DERs to track the dispatch order and provide ancillary services to the system operator. CVPP is responsible for trading deal in the electricity markets, including the transaction price and amount of electricity. It helps the small-scale DERs to participate in the electricity market. Usually, a TVPP has one point of common coupling (PCC). On the contrary, a CVPP may consist of multiple TVPPs belonging to the same entity. In this paper, we focus on the technical power flexibility of a VPP, so our research object is TVPP.

In a VPP, the DERs are aggregated together. Evaluating power flexibility of VPP is critical to participate in power system operation or electricity market. Some literatures have proposed methods to assess the deterministic power flexibility range [4]–[6]. However, there are many random factors that can affect the flexibility range of VPP, such as the loads' variations and the fluctuations of renewable generators' output. The evaluation of VPP's flexibility should reflect the characteristics of these uncertainties. i.e., it should be formulated as a stochastic model.

In this paper, we propose a model to assess the flexibility of VPP considering the randomness of DERs. In this model, we focus on the active and reactive power adjustable capability of the whole VPP at PCC. The randomness of the model is described as the confidence that the operational constraints are satisfied. The stochastic flexibility model can help the VPP operators to evaluate voltage regulation and power regulation capabilities. With the chance constraint or two-stage stochastic programming method [7], this model can be easily incorporated in the stochastic unit commitment [8], risk dispatching [9], biding in the electricity market [10] and so on.

*B. Related work and contribution*

The aggregation of power flexibility problem has been studied in several previous works. The power capability diagram is used in [5] to describe the power capability of microgrid and the impacts of plug-in hybrid electric vehicles (PHEVs), capacitor banks and storage devices are discussed. Similarly, the power flexibility of DERs in different scenarios and degrees of control is introduced in [11]. Silva et al developed the methodology [12] to find the flexibility area at the Transmission/Distribution System Operator (TSO/DSO) boundary node. Then, they proposed the Interval Constrained Power Flow (ICPF) method to estimate the flexibility and corresponding adjustment costs [13]. In [14], the grid scanning method is used to obtain the maximum flexibility potential of the active distribution grid. The work of [15] uses the Monte-Carlo method to assess the probabilistic Interconnection Power Flow Feasible Operation Region (IPF-FOR) for a future time interval by testing a large number of random forecast error scenarios. The linear models of flexibility ranges of units and linear power flow model are used in [16] to reduce the computation time while maintaining accuracy. Reference [17] proposed the linear OPF-based flexibility aggregation algorithm and use beta distribution to consider the random variables. It can be conveniently used in the large grid models with a big amount of flexibility resources. Besides, the effect of different constraints on the shape of the power capability chart is discussed in [17]–[19].

In this paper, we propose a method to aggregate the power flexibility ranges of the DERs in the VPP. As shown in Fig. 1, each kind of DER has a power flexibility range, which can be expressed as a range on the P-Q panel. Besides, there are some time-coupling constraints such as the ramping constraints of generators and energy constraints of energy storage systems (ESS). The power flexibility ranges and time-coupling constraints constitute the technical constraints of DERs. Besides, the network constraints should be considered, such as the voltage limit constraints of buses and the capacity constraints of transmission lines. The main goal of our method is to aggregate the power flexibility ranges of DERs and assess the power flexibility at the PCC while meeting these constraints. However, the loads and maximum output power of renewable generators are random variables, which can be described by the probability distribution functions. Therefore, the performance of the power flexibility range at the VPP's point of common coupling is also stochastic. Hereafter, it is named as the stochastic power flexibility range (PFR) of VPP. Moreover, considering the time-coupling characteristics of serval DERs, a combined battery and generator like model is developed to describe the time-coupling flexibility (TCF) of VPP. By incorporating the stochastic PFR and TCF together, the stochastic flexibility of VPP can be evaluated.

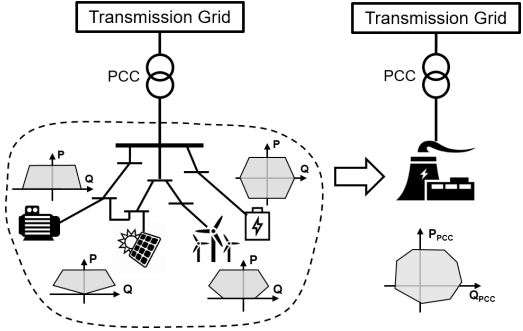

Fig. 1. The schematic diagram of assessing the stochastic power flexibility range (stochastic PFR) of VPP

To the best knowledge of the authors, the main contributions of this paper are listed as follows:

1) We propose a novel stochastic power flexibility evaluation model of VPP, in which Gaussian mixture model (GMM) is used to characterize the correlation between DERs and probability distribution of their power output. GMM can accurately model an arbitrary probability density function (PDF) [20], so it is a more powerful and accurate tool to capture the diverse probability characteristics of DERs as well as their correlations.

2) Precise linearization techniques are used to model the network and DERs, and chance constrained optimization (CCOP) model is developed. The stochastic constraints are transformed into the equivalent deterministic constraints resorting to the affine invariance of GMM. Finally, this problem is converted to a convex optimization problem which can be solved efficiently.

3) A data-driven method is proposed to obtain the piecewise-linear cost function of VPP, which aggregates the cost models of all DERs and can be easily embedded into the optimization model for the power system operator or market bidding.

The remainder of this paper is organized as follows. Section II introduces the network and DERs models. In Section III, the solution procedure of stochastic PFR and TCF are presented. Then, Section IV develops a method to calculate the aggregated cost function of VPP. In Section V, we use a numerical case to demonstrate the application of our proposed method and compare the performance of the existing algorithms. Finally, conclusions are drawn, and further discussions are presented in Section VI.

## II. NETWORK AND DER MODEL

### A. Network model

In VPPs, their distribution networks are usually multi-phase unbalanced. The DERs in the network can be wye-connection, delta-connection, or the combination of the two. In this paper, we use a multi-phase unbalanced network model [21], [22] to simulate the real situations. This model considers all the connection types above and linearizes the multi-phase unbalanced power flow. The complex injection power of wye- and delta-connection sources at bus $i$ can be denoted as the vectors $s_i^Y := [s_i^a, s_i^b, s_i^c]^\top$ and $s_i^\Delta := [s_i^{ab}, s_i^{bc}, s_i^{ca}]^\top$, respectively. And their collections are defined as

$$s^Y := \left((s_1^Y)^\top, (s_2^Y)^\top, \cdots, (s_N^Y)^\top\right)^\top = p^Y + \jmath q^Y \quad (1)$$

$$s^\Delta := \left((s_1^\Delta)^\top, (s_2^\Delta)^\top, \cdots, (s_N^\Delta)^\top\right)^\top = p^\Delta + \jmath q^\Delta \quad (2)$$

Here, $p^Y, p^\Delta, q^Y, q^\Delta$ represent the active and reactive injection power in different connection forms. Finally, we arrange all the injection power as

$$x := \left[(p^Y)^\top, (q^Y)^\top, (p^\Delta)^\top, (q^\Delta)^\top\right]^\top \quad (3)$$

Then, the voltage amplitudes $V$, complex power injection $s_0$ at PCC, and branch current $i_{ij}$ can be expressed as the following linear form, respectively.

$$V = Kx + b \quad (4)$$

$$i_{ij} = Jx + d_{ij} \quad (5)$$

$$s_0 = g^\top x + c \quad (6)$$

Where, the matrices $K, J$, vectors $g, b, d_{ij}$ and constant $c$, are all the system parameters, whose detail definitions can be referred in [21], [22]. This linearization method is essentially a linearized interpolation of the two load-flow solutions: the given operation point and a known zero-load operation point. This multiphase linear model also has a good approximation accuracy performance. According to the numerical tests results of [21], the relative errors of voltages amplitude in the IEEE 13 system and a real system with about 2000 nodes are less than 0.2% and 0.6%, respectively.

### B. DER model

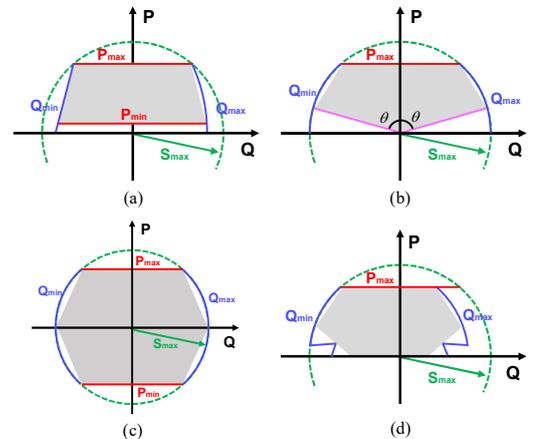

Fig. 2. The power regulation capability charts of (a) combined heat and power generator; (b) inverter-interfaced generator; (c) energy storage battery; and (d) doubly-fed induction generator. The grey polygons represent the linearized regulation capability.

The regulation capability charts of DERs' outputs have different shapes [4]. In this section, we formulate following different DERs including the combined heat and power generator (CHP), inverter-interfaced generator (such as PV), energy storage system (ESS) and doubly-fed induction generator (DFIG) [23]. Their regulation capability charts consist of the constraints associated with DERs' parameters, which usually include non-linear or even non-convex constraints. Therefore, we linearize their boundaries and convert these charts into polygons as shown in Fig. 2.

The polygonal power regulation capability charts of these DERs can be described as the following polygons:

$$A_{i,s}^\psi \cdot \begin{bmatrix} P_{i,s}^{\psi,t} \\ Q_{i,s}^{\psi,t} \end{bmatrix} \leq b_{i,s}^\psi \quad (7)$$

$$s \in \{\text{CHP,PV,ESS,WT}\}$$

Where, these linear constraints represent the feasible operation region of the DERs, depicted as the gray polygonal areas in Fig. 2. For example, $P_{i,\text{CHP}}^{\psi,t}$ denotes the active output power of CHP on the phase $\psi$ of bus $i$ at time $t$. The superscript $\psi$ denotes the connection phases of the DERs. If a CHP is delta-connection on the $ab$ and $bc$ phase, then $\psi = \{ab, bc\}$. By summing the output power of all the phases of one device, the total output active power of this device can be described as follows:

$$P_{i,s}^t = \sum_\psi P_{i,s}^{\psi,t} \quad (8)$$

$$s \in \{\text{CHP,PV,ESS,WT}\}$$

The time-coupled operation constraints of DERs include the ramping constraints of generators (9) and the energy constraints of ESS (10)-(11), which are shown as follows, respectively.

$$-r_{i,\text{CHP}} \leq P_{i,\text{CHP}}^t - P_{i,\text{CHP}}^{t-1} \leq r_{i,\text{CHP}} \quad (9)$$

$$E_{i,\text{ESS}}^t = \alpha_{i,\text{ESS}} E_{i,\text{ESS}}^{t-1} - \Delta t \cdot P_{i,\text{ESS}}^t \quad (10)$$

$$E_{i,\text{ESS}}^{\min} \leq E_{i,\text{ESS}}^t \leq E_{i,\text{ESS}}^{\max} \quad (11)$$

The maximum output power of renewable energy, such as photovoltaics and wind turbines, depend on the weather condition, which are random factors. Their forecast output power can be described as the sum of expected forecast value ($\bar{P}_{i,\text{PV}}^{t,\max}$, $\bar{P}_{i,\text{WT}}^{t,\max}$) and forecast errors ($\tilde{e}_{i,\text{PV}}^t$, $\tilde{e}_{i,\text{WT}}^t$). Curtailed power $\hat{P}_{i,\text{PV}}^t$ and $\hat{P}_{i,\text{WT}}^t$ are used to provide flexibility.

$$\begin{aligned} \tilde{P}_{i,s}^{t,\max} &= \bar{P}_{i,s}^{t,\max} + \tilde{e}_{i,s}^t \\ P_{i,s}^t &= \bar{P}_{i,s}^{t,\max} - \hat{P}_{i,s}^t \\ 0 &\leq \hat{P}_{i,s}^t \leq \bar{P}_{i,s}^{t,\max} \end{aligned} \quad (12)$$

$$s \in \{\text{PV,WT}\}$$

The load in VPP depends on the behavior of consumers, and they are also random variables. We can assume that their power factor is constant [16], that is:

$$\tilde{P}_{i,\text{load}}^{\psi,t} = \bar{P}_{i,\text{load}}^{\psi,t} + \tilde{e}_{i,\text{load}}^{\psi,t} \quad (13)$$

$$\tilde{Q}_{i,\text{load}}^{\psi,t} = \tilde{P}_{i,\text{load}}^{\psi,t} \tan\left(\varphi_{i,\text{load}}^\psi\right) \quad (14)$$

By summing the power DERs and loads, we can get the power injection of buses.

$$\tilde{P}_{i,\text{inj}}^{\psi,t} = P_{i,\text{CHP}}^{\psi,t} + P_{i,\text{PV}}^{\psi,t} + P_{i,\text{ESS}}^{\psi,t} + P_{i,\text{WT}}^{\psi,t} - \tilde{P}_{i,\text{load}}^{\psi,t} \quad (15)$$

$$\tilde{Q}_{i,\text{inj}}^{\psi,t} = Q_{i,\text{CHP}}^{\psi,t} + Q_{i,\text{PV}}^{\psi,t} + Q_{i,\text{ESS}}^{\psi,t} + Q_{i,\text{WT}}^{\psi,t} - \tilde{Q}_{i,\text{load}}^{\psi,t} \quad (16)$$

Accordingly, the power injection in the wye- and delta-connection format at time $t$ is given by

$$p^{Y,t} = \left[\tilde{P}_{i,\text{inj}}^{\psi,t}\right]_{i \in \mathcal{N}_Y, \psi \subseteq Y} \quad (17)$$

$$q^{Y,t} = \left[\tilde{Q}_{i,\text{inj}}^{\psi,t}\right]_{i \in \mathcal{N}_Y, \psi \subseteq Y} \quad (18)$$

$$p^{\Delta,t} = \left[\tilde{P}_{i,\text{inj}}^{\psi,t}\right]_{i \in \mathcal{N}_\Delta, \psi \subseteq \Delta} \quad (19)$$

$$q^{\Delta,t} = \left[\tilde{Q}_{i,\text{inj}}^{\psi,t}\right]_{i \in \mathcal{N}_\Delta, \psi \subseteq \Delta} \quad (20)$$

Based on the multi-phase unbalanced network model, we can calculate buses' voltages and the branches' currents. Considering the randomness of renewable energy generators and loads, the network operational constraints are formulated as chance constraints as follows:

$$\Pr\left\{\tilde{V}_i^{\psi,t} \leq V_i^{\psi,\max}\right\} \geq 1 - \alpha^{V+} \quad (21)$$

$$\Pr\left\{\tilde{V}_i^{\psi,t} \geq V_i^{\psi,\min}\right\} \geq 1 - \alpha^{V-} \quad (22)$$

$$\Pr\left\{\tilde{I}_{ij}^{\psi,t} \leq I_{ij}^{\psi,\max}\right\} \geq 1 - \alpha^{I+} \quad (23)$$

$$\Pr\left\{\tilde{I}_{ij}^{\psi,t} \geq -I_{ij}^{\psi,\max}\right\} \geq 1 - \alpha^{I-} \quad (24)$$

Due to the volatility of loads and renewable energy generators, the random capability of maximum output active power at PCC, denoted as $\tilde{P}_{\text{rand}}^{t,\max}$, is also a random variable.

$$\tilde{P}_{\text{rand}}^{t,\max} = \sum_i \sum_\psi \left(\tilde{P}_{i,\text{PV}}^{\psi,t,\max} + \tilde{P}_{i,\text{WT}}^{\psi,t,\max} - \tilde{P}_{i,\text{load}}^{\psi,t}\right) \quad (25)$$

Then, the following chance constraint can be used to express the influence of volatility on the output active power capability.

$$\Pr\left\{P_{\text{rand}}^t \leq \tilde{P}_{\text{rand}}^{t,\max}\right\} \geq 1 - \alpha^P \quad (26)$$

Where, $P_{\text{rand}}^t$ denotes the actual capability of maximum output active power at PCC, as shown in (27):

$$P_{\text{rand}}^t = \sum_i \sum_\psi \left(P_{i,\text{PV}}^{\psi,t} + P_{i,\text{WT}}^{\psi,t} - \bar{P}_{i,\text{load}}^{\psi,t}\right) \quad (27)$$

### III. SOLUTION PROCEDURE

#### A. Decision variables

To simplify the expression of decision variables, we use the vectors to collect all the output power of DERs at time $t$:

$$P_s^t = \left[P_{i,s}^{\psi,t}\right]_{\forall i, \forall \psi}, Q_s^t = \left[Q_{i,s}^{\psi,t}\right]_{\forall i, \forall \psi} \quad (28)$$

$$s \in \{\text{SG,PV,ESS,WT}\}$$

Then, we use $x^t$ to represent the decision variables vector at time $t$, which is made up of all the output power of controllable DERs:

$$x^t = \begin{bmatrix} P_{\text{CHP}}^t; Q_{\text{CHP}}^t; P_{\text{PV}}^t; Q_{\text{PV}}^t; \\ P_{\text{ESS}}^t; Q_{\text{ESS}}^t; P_{\text{WT}}^t; Q_{\text{WT}}^t \end{bmatrix} \quad (29)$$

#### B. Modeling of uncertainties

The forecast errors of renewable energy generators and loads constitute the random variables. We use vector $\tilde{e}^t$ to denote the collection of them at time $t$.

$$\tilde{e}^t = \left[\tilde{e}_{i,\text{PV}}^{\psi,t}; \tilde{e}_{i,\text{WT}}^{\psi,t}; \tilde{e}_{i,\text{load}}^{\psi,t}\right]_{\forall i, \forall \psi} \quad (30)$$

The joint probability density function (PDF) of $\tilde{e}^t$ can be estimated from the historical data. In this paper, we use Gaussian Mixture Model (GMM) to characterize uncertainties of forecast errors. GMM can be used to fit any PDF of random variables by adjusting its parameters and keeps affine invariance [8]. With GMM, $\tilde{e}^t$ can be expressed by an affine combination of multivariate Gaussian distributions as follows:

$$PDF_{\tilde{e}^t}(x) = \sum_{j=1}^{n} \omega_j^t \mathcal{N}\left(x; \mu_j^t, \Sigma_j^t\right)$$
$$\sum_{j=1}^{n} \omega_j^t = 1, \omega_j^t > 0 \quad (31)$$

Where, $\omega_j^t$ is the weight coefficient; $\mu_j^t$ is the expectation vector; and $\Sigma_j^t$ is the covariance matrix of $j$-th Gaussian distribution vector.

### C. Conversion of chance constraints

In our model, the network operational constraints are formulated as chance constraints. To make them solvable, they should be converted to deterministic ones. We take the chance constraints (21)-(22) related $\tilde{V}_i^{\psi,t}$ as the example to demonstrate the solution.

Since the network model (4)-(5) is linear, the bus voltages and branch currents can be expressed in the affine form of forecast errors and decision variables.

$$\tilde{V}_i^{\psi,t} = \left(a_{V,i}^\psi\right)^\top \tilde{e}^t + \left(b_{V,i}^\psi\right)^\top x^t + c_{V,i}^\psi \quad (32)$$

$$\tilde{I}_{ij}^{\psi,t} = \left(a_{I,ij}^\psi\right)^\top \tilde{e}^t + \left(b_{I,ij}^\psi\right)^\top x^t + c_{I,ij}^\psi \quad (33)$$

$$\tilde{P}_{\text{rand}}^{t,\max} = \left(a_P\right)^\top \tilde{e}^t + c_{P1} \quad (34)$$

$$P_{\text{rand}}^t = \left(b_P\right)^\top x^t + c_{P2} \quad (35)$$

Where, $a_{V,i}^\psi$, $b_{V,i}^\psi$, $a_{I,ij}^\psi$, $b_{I,ij}^\psi$, $a_P$ and $b_P$ are constant coefficient vectors, $c_{V,i}^\psi$, $c_{I,ij}^\psi$, $c_{P1}$ and $c_{P2}$ are constant coefficients. They can be derived from the linear network model.

Since GMM keeps affine invariance (that means the affine combination of GMM is still a GMM), the PDF of single variable $(a_{V,i}^\psi)^\top \tilde{e}^t$ can be expressed as follows:

$$PDF_{(a_{V,i}^\psi)^\top \tilde{e}^t}(x) = \sum_{j=1}^{n} \omega_j^t \mathcal{N}\left(x; \mu_{V,i}^{\psi,t}, \left(\sigma_{V,i}^{\psi,t}\right)^2\right) \quad (36)$$

Where, the expectations and variances of PDF are:

$$\mu_{V,i}^{\psi,t} = \left(a_{V,i}^\psi\right)^\top \mu_j^t$$
$$\left(\sigma_{V,i}^{\psi,t}\right)^2 = \left(a_{V,i}^\psi\right)^\top \Sigma_j^t \left(a_{V,i}^\psi\right) \quad (37)$$

Further, the cumulative density function (CDF) of variable $(a_{V,i}^\psi)^\top \tilde{e}^t$ can be transformed into the linear combination of standard Gaussian distributions, where $\Phi(x)$ denotes CDF of standard Gaussian distribution of variable $x$.

$$CDF_{(a_{V,i}^\psi)^\top \tilde{e}^t}(x) = \sum_{j=1}^{n} \omega_j^t \Phi\left(\frac{x - \mu_{V,i}^{\psi,t}}{\sigma_{V,i}^{\psi,t}}\right) \quad (38)$$

Therefore, the chance constraints (21)-(22) can be transformed into the equivalent deterministic constraints with the quantiles of PDF.

$$V_i^{\psi,\max} - \left(b_{V,i}^\psi\right)^\top x^t - c_{V,i}^\psi \geq Quant\left(1-\alpha^{V+} \left| \left(a_{V,i}^\psi\right)^\top \tilde{e}^t\right.\right) \quad (39)$$

$$V_i^{\psi,\min} - \left(b_{V,i}^\psi\right)^\top x^t - c_{V,i}^\psi \leq Quant\left(\alpha^{V-} \left| \left(a_{V,i}^\psi\right)^\top \tilde{e}^t\right.\right) \quad (40)$$

Similarly, chance constraints (23)-(24) related to variable $\tilde{I}_i^{\psi,t}$ and chance constraint (26) can be transformed into

$$I_{ij}^{\psi,\max} - \left(b_{I,ij}^\psi\right)^\top x^t - c_{I,ij}^\psi \geq Quant\left(1-\alpha^{I+} \left| \left(a_{I,ij}^\psi\right)^\top \tilde{e}^t\right.\right) \quad (41)$$

$$-I_{ij}^{\psi,\max} - \left(b_{I,ij}^\psi\right)^\top x^t - c_{I,ij}^\psi \leq Quant\left(\alpha^{I-} \left| \left(a_{I,ij}^\psi\right)^\top \tilde{e}^t\right.\right) \quad (42)$$

$$\left(b_P\right)^\top x^t - c_{P1} + c_{P2} \leq Quant\left(\alpha^P \left| \left(a_P\right)^\top \tilde{e}^t\right.\right) \quad (43)$$

Where, the quantiles are defined as the inverse of cumulative density function (CDF):

$$Quant(\alpha | \tilde{x}) \triangleq CDF_{\tilde{x}}^{-1}(\alpha) \quad (44)$$

For example, if $Quant(\alpha | \tilde{x}) = q$ or $\alpha = CDF_{\tilde{x}}(q)$. The value of $q$ can be solved iteratively with Newton method as the pseudo-code in Algorithm 1.

| **Algorithm 1.** Calculate Quantile with Newton Method |
|---|
| 1: Given the initial value: $q_i = q_0, i = 0$, maximum error $\varepsilon$ |
| 2: **LOOP UNTIL** $\left\|CDF_{\tilde{x}}(q_i) - \alpha\right\| < \varepsilon$ |
| 3: $q_{i+1} = q_i - \dfrac{CDF_{\tilde{x}}(q_i) - \alpha}{PDF_{\tilde{x}}(q_i)}$, $i \leftarrow i+1$ |
| 4: **END LOOP** |

### D. Solution of stochastic power flexibility range

#### 1) OPF model for PFR

Based on the multi-phase linear network model (6), we can calculate the maximal complex power injection of PCC at each time $t$, denoted by $P_{\text{PCC}}^t$ and $Q_{\text{PCC}}^t$.

$$P_{\text{PCC}}^t = \left(m^t\right)^\top x^t + g^t \quad (45)$$

$$Q_{\text{PCC}}^t = \left(h^t\right)^\top x^t + l^t \quad (46)$$

Specifically, for given the confidence $\gamma = 1-\alpha$ for chance constraints (21)-(24), (26) and the power factor $\varphi$ of PCC, an optimal power flow (OPF) is developed to calculate the power flexibility range:

The objective function is to maximize the regulation power capability of VPP under a given power factor $\varphi$.

$$\max_{X^t} P_{\text{PCC}}^t \cdot \cos(\varphi) + Q_{\text{PCC}}^t \cdot \sin(\varphi) \quad (47)$$

The constraints include four parts:
(i) network model (45)-(46) and

$$Q_{\text{PCC}}^t = P_{\text{PCC}}^t \cdot \tan(\varphi) \quad (48)$$

(ii) the constraints of time-decoupling DERs (7);
(iii) maximum output power of renewable energy constraints (12);
(iv) the equivalent deterministic constraints (39)-(43).

Here, the power factor $\varphi$ is a parameter. By varying $\varphi$ from 0 to $2\pi$ [18], a series of OPFs are conducted to get a group of results, forming the power flexibility range with the confidence level $\gamma$. As shown in Fig. 3, the power flexibility range varying with $\varphi$ can be obtained, which represents the maximal regulation power capability.

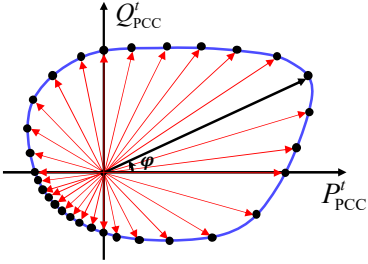

Fig. 3. The PFR of VPP with power factor $\varphi$ and specific confidence level $\gamma$.

If the confidence level changes, the values of quantiles in the network constraints (39)-(42) change accordingly. Then a new PFR is obtained under the new confidence level. A function can be constructed, which maps from the PFR to the confidence.

$$\gamma = Conf_t\left(P_{\mathrm{PCC}}^t, Q_{\mathrm{PCC}}^t\right) \quad (49)$$

In the 3-dimensional space of $P_{\mathrm{PCC}}^t$, $Q_{\mathrm{PCC}}^t$ and $\gamma$, this function is in the form of a surface. Each group of results under the same confidence forms a contour of this surface. By solving the OPFs with different confidence, the points on the corresponding contours can be calculated. Furthermore, all the points on the surface can be obtained, which is the stochastic PFR of the VPP.

*2) Analytic reformulation of the stochastic PFR*

In the section above, the stochastic PFR of VPP is obtained, which is a surface composed of a large number of scattered points. Nevertheless, an analytic expression is preferable to describe the function (49). Here, we use the convex piecewise-linear fitting algorithm [24] to fit this function, whose process is explained in detail in the supplementary file [25]. Therefore, the analytic reformulation of the stochastic PFR can be expressed as the piecewise-linear function with $m$ partitions:

$$Conf_t\left(P_{\mathrm{PCC}}^t, Q_{\mathrm{PCC}}^t\right) = \mathrm{ReLU}\left(\min_{j=1,\ldots,m}\left\{\boldsymbol{a}_j^\top \begin{bmatrix} P_{\mathrm{PCC}}^t \\ Q_{\mathrm{PCC}}^t \end{bmatrix} + b_j\right\}\right) \quad (50)$$

Where, ReLU(•) denotes the Rectified Linear Unit function. Since $\gamma$ is confidence level and it cannot be negative, we just need to fit the positive part of the function. The convex piecewise-linear fitting algorithm can be applied in the fitting process, because the positive part of this function is approximately concave.

Therefore, given a specific confidence level $\gamma$, we can calculate the corresponding PFR at time $t$ based on (50). It can be expressed as a polygon:

$$\mathcal{R}_\gamma^t = \left\{\left(P_{\mathrm{PCC}}^t, Q_{\mathrm{PCC}}^t\right) \middle| \boldsymbol{A}_\gamma^t \begin{bmatrix} P_{\mathrm{PCC}}^t \\ Q_{\mathrm{PCC}}^t \end{bmatrix} \leq \boldsymbol{b}_\gamma^t\right\} \quad (51)$$

Where, $\boldsymbol{A}_\gamma^t$ and $\boldsymbol{b}_\gamma^t$ are the constant coefficients matrix and vector of this polygon, respectively.

*E. Solution of time-coupling flexibility*

The PFR of VPP reflects the time-decoupling constraints of DERs in each period. However, there are also time-coupling constraints, such as (9)-(11), which affects the active output power of VPP among different periods. Therefore, extra constraints are needed to describe the time-coupling flexibility (TCF) of VPP.

Given a specific confidence level $\gamma$, the TCF can be described as a combined battery and generator like model, whose key parameters include: (i) maximum and minimum active output power at every period $t$ (e.g.: $P_{\mathrm{PCC},\gamma}^{t,\max}$, $P_{\mathrm{PCC},\gamma}^{t,\min}$); (ii) maximum up and down ramping rates between any two adjacent time points $t$ and $t+1$ (e.g.: $r_{\mathrm{PCC},\gamma}^{t,\mathrm{up}}$, $r_{\mathrm{PCC},\gamma}^{t,\mathrm{down}}$); and (iii) maximum and minimum energy constraints in the first $t$ periods (e.g.: $E_{\mathrm{PCC},\gamma}^{t,\max}$, $E_{\mathrm{PCC},\gamma}^{t,\min}$). Then, the flexibility region of active output power of VPP among different periods can be described as

$$\begin{aligned}
\mathcal{C}_\gamma = \{\boldsymbol{P}_{\mathrm{PCC}} \mid &P_{\mathrm{PCC},\gamma}^{t,\min} \leq P_{\mathrm{PCC}}^t \leq P_{\mathrm{PCC},\gamma}^{t,\max}, \\
&r_{\mathrm{PCC},\gamma}^{t,\mathrm{down}} \leq P_{\mathrm{PCC}}^{t+1} - P_{\mathrm{PCC}}^t \leq r_{\mathrm{PCC},\gamma}^{t,\mathrm{up}}, \\
&E_{\mathrm{PCC},\gamma}^{t,\min} \leq \sum_{\tau=1}^{t} P_{\mathrm{PCC}}^\tau \Delta t \leq E_{\mathrm{PCC},\gamma}^{t,\max}, \forall t \in \mathcal{T}\}
\end{aligned} \quad (52)$$

The value of these key parameters of TCF should guarantee the disaggregation feasibility [26]. That is to say, given any $\boldsymbol{P}_{\mathrm{PCC}} \in \mathcal{C}_\gamma$, there must exist a disaggregation solution $\boldsymbol{x}^*$, such that $\boldsymbol{P}_{\mathrm{PCC}}$ can be realized according to (45). Based on the robust modification method proposed in [27], the key parameters of TCF can be solved in a heuristic approach, which is a two-stage method.

---

**Algorithm 2.** Robust modification method

1: Initialize the value of $P_{\mathrm{PCC},\gamma}^{t,\max}$, $P_{\mathrm{PCC},\gamma}^{t,\min}$, $r_{\mathrm{PCC},\gamma}^{t,\mathrm{up}}$, $r_{\mathrm{PCC},\gamma}^{t,\mathrm{down}}$, $E_{\mathrm{PCC},\gamma}^{t,\max}$, and $E_{\mathrm{PCC},\gamma}^{t,\min}$ with the extreme points. Given the reduction ratio $\theta$ and the maximum tolerance of violations $\varepsilon$.
2: Solve the max-min problem (53)-(55), obtain the optimal objective value $f^*$.
3: If $|f^*| < \varepsilon$, feasible key parameters obtained, otherwise modify the parameters according to (56) and go to step 2.

---

In the first stage, we test the disaggregation feasibility of the latest parameters. The maximum deviation of the output active output power at PCC can be calculated by solving the following max-min problem

$$f = \max_{\boldsymbol{P}_{\mathrm{PCC}}} \min_{\boldsymbol{x}} \sum_{t=1}^{T}\left(P_{\mathrm{PCC}}^{t,+} + P_{\mathrm{PCC}}^{t,-}\right) \quad (53)$$

$$s.t. \quad (7), (9)\text{-}(12), (39)\text{-}(43) \quad \forall t \in \mathcal{T}$$

$$P_{\mathrm{PCC}}^t = \left(\boldsymbol{m}^t\right)^\top \boldsymbol{x}^t + g^t + P_{\mathrm{PCC}}^{t,+} - P_{\mathrm{PCC}}^{t,-}, \forall t \in \mathcal{T} \quad (54)$$

$$P_{\mathrm{PCC}}^{t,+} \geq 0, P_{\mathrm{PCC}}^{t,-} \geq 0, \forall t \in \mathcal{T} \quad (55)$$

Where, $P_{\mathrm{PCC}}^{t,+}$ and $P_{\mathrm{PCC}}^{t,-}$ denote the positive and negative violations. If the maximum violations do not equal to 0, it means that the current value of these parameters are not disaggregation feasible, and further modification is needed.

Then, in the second stage, all the parameters are modified as (56) to narrow the flexibility region $\mathcal{F}_\gamma$, so that the violations will be reduced in the next iteration of the first stage.

$$\pi_{(k+1)}^{\min} = \theta \pi_{(k)}^{\min} + (1-\theta) \pi_{(k)}^{\max}$$
$$\pi_{(k+1)}^{\max} = \theta \pi_{(k)}^{\max} + (1-\theta) \pi_{(k)}^{\min} \quad (56)$$

Where, $\pi_{(k)}^{\min}$ denotes $P_{PCC,\gamma}^{t,\min}$, $r_{PCC,\gamma}^{t,down}$, or $E_{PCC,\gamma}^{t,\min}$ for any $t \in \mathcal{T}$ at the $k$-th iteration; $\pi_{(k)}^{\max}$ denotes $P_{PCC,\gamma}^{t,\max}$, $r_{PCC,\gamma}^{t,up}$, or $E_{PCC,\gamma}^{t,\max}$ for any $t \in \mathcal{T}$ at the $k$-th iteration, correspondingly; $\theta$ is a constant representing the reduction ratio of each step.

The robust modification method is presented as Algorithm 2.

*F. Integration of flexibility model and application in the stochastic unit commitment framework*

By integrating the stochastic PFR at every period of time and TCF of VPP, we can finally obtain the whole flexibility model. Given a specific confidence level $\gamma$, the stochastic flexibility model of VPP can be described as

$$\mathcal{F}_\gamma = \left\{ (\boldsymbol{P}_{PCC}, \boldsymbol{Q}_{PCC}) \mid (P_{PCC}^t, Q_{PCC}^t) \in \mathcal{R}_\gamma^t, \forall t \in \mathcal{T}, \boldsymbol{P}_{PCC} \in \mathcal{C}_\gamma \right\} \quad (57)$$

The flexibility model of VPP can be applied to the stochastic unit commitment framework with VPPs, conventional thermal generation units and renewable power plants. Its objective is to minimize the total cost, including the VPP cost, start up and shut down cost, fuel cost, reserve cost and the potential curtailment penalty of renewable energy. The constraints include power balance constraints, generator constraints, renewable energy constraints, VPP stochastic flexibility constraints, system reserve chance constraints, and network chance constraints. The detailed stochastic unit commitment model can refer to [8]. Then, the DERs in the VPP can participate in the stochastic unit commitment as a special power plant.

## IV. CALCULATING THE COST FUNCTION OF VPP

To participate in system operation or market bidding, the VPP need calculate its aggregating cost function. This section will introduce the solution for generating the cost function of VPP.

*A. Cost model of DERs*

Firstly, we construct the cost function model of each DER in the VPP. The cost functions of CHPs are quadratic:

$$C_{i,CHP}(P_{i,CHP}^t) = a_{i,CHP}(P_{i,CHP}^t)^2 + b_{i,CHP}(P_{i,CHP}^t) + c_{i,CHP} \quad (58)$$

The operation and maintenance cost of ESS is as follows, with $K_i^{ch}$ and $K_i^{dis}$ denote the charge and discharge cost parameters of ESS, respectively.

$$C_{i,ESS}(P_{i,ESS}^t) = \max \left\{ K_i^{dis} P_{i,ESS}^t \Delta t, -K_i^{ch} P_{i,ESS}^t \Delta t \right\} \quad (59)$$

Where, $\Delta t$ denotes the time interval of one period.

The marginal generation costs of PV and WT in the VPP are negligible and we set them to 0.

*B. Piecewise-linear fitting the cost function of VPP*

Using the stochastic PFR model presented in Section III, we can obtain the minimum and maximum active output power of VPP at time $t$, denoted by $P_{PCC}^{t,\min}$ and $P_{PCC}^{t,\max}$. Then, we can sample a series of operation points equally in interval $\left[ P_{PCC}^{t,\min}, P_{PCC}^{t,\max} \right]$, and solve the flowing OPF to find the minimum operating cost of VPP, corresponding to the given $P_{PCC}^t$.

$$\min_{X^t} C_{VPP}^{t,(k)} = \sum_{i=1}^N \begin{pmatrix} C_{i,DG}(P_{i,DG}^t) + C_{i,wind}(P_{i,wind}^t) \\ + C_{i,PV}(P_{i,PV}^t) + C_{i,ESS}(P_{i,ESS}^t) \end{pmatrix} \quad (60)$$

$$s.t. \quad (7), (12), (39)\text{-}(43), (45)$$
$$P_{PCC}^t = P_{PCC}^{t,(k)} \quad (61)$$

Here, $C_{VPP}^{t,(k)}$ denotes the minimum cost of VPP corresponding to the $k$-th sample $P_{PCC}^{t,(k)}$. Therefore, every tuple $(P_{PCC}^{t,(k)}, C_{VPP}^{t,(k)})$ corresponds to a point on the cost function curve of VPP. In this way, we can get $K$ sample points in total. With the convex piecewise-linear fitting algorithm in the supplementary file [25], these samples can be further used to fit the analytic expression of the cost function into $m$ linear partitions and can be easily incorporated to any optimization problem as follows:

$$\min \quad Cost_t(P_{PCC}^t) = r$$
$$s.t. \quad a_j^T P_{PCC}^t + b_j \leq r, j=1,\ldots,m \quad (62)$$

## V. NUMERICAL TESTS

*A. Simulation setup*

Numerical tests are carried out on the 15-bus modified European medium voltage distribution network benchmark [28]. The topology of network and detail parameters of DER units can refer to the supplementary file [25]. The maximum and minimum limits of buses' voltage are set to 1.05 p.u. and 0.95 p.u., respectively. The output power data of PVs, wind turbines and loads are cited from [29] and [30]. These historical data are used to generate the probability density function of forecast errors using GMM.

The test case was conducted on a laptop with Intel Core i7-8550U CPU, 1.80 GHz and 16 GB RAM. The MATLAB software with YALMIP toolbox and CPLEX solver were used to solve the optimization problems.

We firstly use the proposed methodology to evaluate the stochastic PFR in section B, with the comparison of computational efficiency is presented. The results of TCF model is shown in Section C. Besides, Section D introduces the aggregated piecewise-linear cost function of VPP. Finally, the application of VPP flexibility model in the stochastic unit commitment is presented Section E.

*B. Stochastic power flexibility range evaluation*

Based on our proposed method, we can obtain the stochastic PFR of VPP. We use the operational state of the VPP at 12:00 as an example. The result is a three-dimensional surface composed of many scattered points. The X-axis and Y-axis represents the injection active and reactive power at PCC of VPP, and the Z-axis represents the confidence level corresponding to the injected power. To express the results analytically, the obtained scatters are fitted into a 16 partitions' piecewise-linear function, as shown in Fig. 4 (a).

Therefore, given a specific confidence level $\gamma$, we can calculate the corresponding PFR based on (50). It can be expressed as a polygon as (51). The size of $A_\gamma^t$ and $b_\gamma^t$ are $(16 \times 2)$ and $(16 \times 1)$, respectively. Set the confidence level as $\gamma=0.8$, the numerical values of $A_\gamma^t$ and $b_\gamma^t$ can refer to the supplementary file [25].

What is more, the stochastic PFR of VPP can also be obtained point by point by the Monte-Carlo simulation method. Based on the expectation vectors and covariance matrices of GMM, we can generate the realization scenarios of the loads and maximum output power of renewable generators. Subsequently, at each realization of injection power at PCC $(P_{\text{PCC}}^{t,(k)}, Q_{\text{PCC}}^{t,(k)})$, we can construct all the operational constraints and find out the proportion of feasible scenarios as the confidence level of this point $(P_{\text{PCC}}^{t,(k)}, Q_{\text{PCC}}^{t,(k)})$. After finishing scanning all the operating points, we can get stochastic PFR of VPP at time $t$. In this test case, we scanned possible injection power points on a 200×200 grid and generated 1,200 scenarios at each point to calculate the confidence level. The results of Monte-Carlo simulation is shown in Fig. 4 (b).

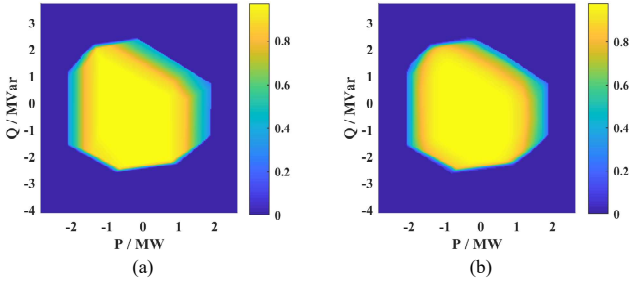

Fig. 4. The stochastic PFR of VPP obtained by (a) piecewise-linear fitting; (b) Monte-Carlo simulation.

In Fig. 4, the warmer is the color, the higher is the confidence level, and the smaller is the corresponding PFR range. This result is reasonable because when the confidence level we required becomes higher, the network constraints becomes stronger, so the PFR result becomes smaller. Besides, the shape of PFR may change when the confidence level changes. That indicates different regions of PFR have different sensitivity to the confidence levels. From TABLE I, we can see the result of Monte-Carlo simulation is similar to that obtained by the proposed method, which verifies the correctness of the proposed method.

TABLE I
ERRORS COMPARISON (RMSE: ROOT MEAN SQUARED ERROR, $R^2$: THE COEFFICIENT OF DETERMINATION)

| Item | RMSE | $R^2$ |
|---|---|---|
| Convex piecewise-linear fitting | 0.0173 | 0.9982 |
| Monte-Carlo simulation | 0.0176 | 0.9979 |

Although the proposed method and the Monte-Carlo simulation method can obtain similar results, the computational efficiency of the two differs greatly. The calculation time and RMSE of our proposed method and the Monte-Carlo simulation are listed in TABLE II. Obviously, the Monte-Carlo simulation cannot be used for real application since of its ultra-heavy computational burden.

TABLE II
COMPUTATIONAL EFFICIENCY COMPARISON

| Item | Calculation time | RMSE |
|---|---|---|
| Our proposed method | 83.19 s | 0.0173 |
| Monte-Carlo simulation | 12,372 s | 0.0176 |

### C. Results of time-coupling flexibility

Based on the solution method of TCP above, the key parameters, including maximum and minimum active output power, maximum up and down ramping rates, and maximum and minimum energy constraints, can be obtained. Taking the confidence level is 0.8 as an example, the result is shown in Fig. 5.

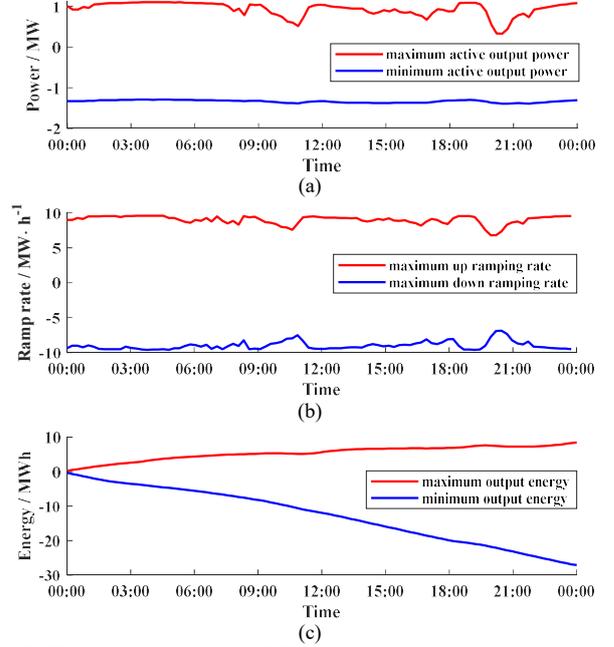

Fig. 5. The key parameters of TCP when the confidence level is 0.8. (a) maximum and minimum active output power; (b) maximum up and down ramping rates; (c) maximum and minimum energy constraints

### D. Aggregated cost function of VPP

With the aggregated cost function generation method, we can solve the piecewise-linear cost function of VPP. The scatters in Fig. 6 shows the original calculation results of the VPP's cost and the polyline is the piecewise-linear fitting result at 12:00.

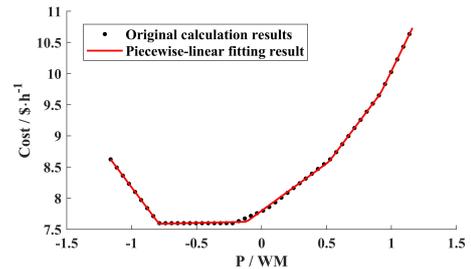

Fig. 6. The cost function of VPP at 12:00.

In this numerical test, the cost function is divided into five partitions. The generation cost function of VPP is adaptively divided into five linear partitions based on the forecast value of loads and output power of renewable energy generators.

Fig. 7 shows the cost function with the granularity of 15 minutes. Because of their fluctuations, the output power range of VPP also changes throughout the day and the cost changes.

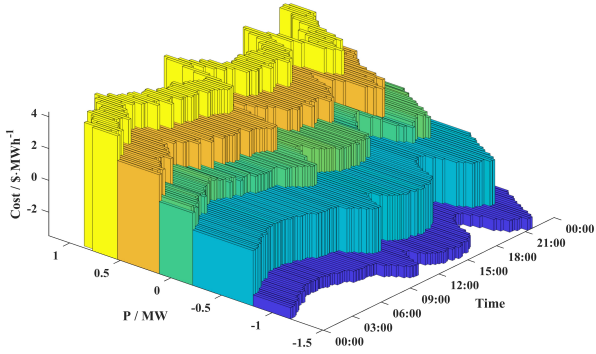

Fig. 7. The aggregated cost of VPP with the granularity of 15 minutes.

We can further calculate the maximum, minimum and average RMSE and $R^2$ of the cost function fitting at different times, as shown in TABLE III.

TABLE III
ERRORS OF THE COST FUNCTION FITTING AT DIFFERENT TIMES (RMSE: ROOT MEAN SQUARED ERROR, $R^2$: THE COEFFICIENT OF DETERMINATION)

| Item | RMSE | $R^2$ |
| --- | --- | --- |
| Maximum | 0.0140 | 0.99999 |
| Minimum | 0.0019 | 0.99974 |
| Average | 0.0099 | 0.99982 |

Combining the stochastic PFR results with the corresponding generation cost functions, the operators of VPPs can further construct their own bidding strategies. They can send the PFR of VPP in the form of a polygon as (57), the segmented nodes of 5 partitions and the corresponding operational cost of VPP to the power system control center. Therefore, the VPPs can participate in the electricity market as a special power plant.

*E. Application of flexibility model in the stochastic unit commitment*

In this section, we demonstrate the application of the VPP flexibility model in the stochastic unit commitment framework. The comparison of two scenarios are used to illustrate the economic value when the VPPs participate in the unit commitment. In Scenario 1, the VPPs are regarded as special power plants. The TSO will first specify a confidence level to the VPP operators. The VPP operators then evaluate the stochastic flexibility of VPPs, including the technical constraints (57) and cost functions (62). Subsequently, the flexibility model will be sent to TSO to participate in the stochastic unit commitment. In Scenario 2, since the operators of VPPs do not have the VPP flexibility evaluation results, they will not provide flexibility to the gird and will not participate in the unit commitment. Each VPP will dispatch the DERs inside to minimize the power purchase at the PCC. In this case, the VPPs operates as normal distribution network and can be regarded as a load.

The IEEE 24-bus test system is used in this numerical test case, which contains 12 thermal generation units, 3 wind farms and 5 VPPs. The parameters of thermal generation units are cited in [31] with partial modification. The flexibility model of each VPP are modified based on the VPP flexibility model above with different parameters of DERs.

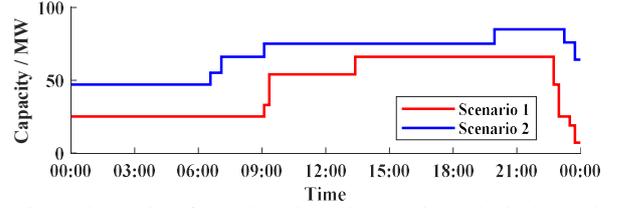

Fig. 8. Total capacity of turned-on thermal generation units in the stochastic unit commitment results.

TABLE IV
TOTAL COST OF TWO SCENARIOS

| Scenarios | Scenario 1 | Scenario 2 |
| --- | --- | --- |
| Total cost ($) | 61,305 | 70,315 |

The results of stochastic unit commitment are shown in Fig. 8 and TABLE IV. With the VPPs evaluation results, the VPP can provide flexibility to the grid and reduce the total operational cost. It achieves the economic benefits through the following three aspects: (i) reduce the total capacity of turned-on thermal generation units; (ii) reduce the reserve needs of thermal generation units; and (iii) provide flexible ramping capacity.

## VI. CONCLUSION

In this paper, we propose a method to assess the stochastic power flexibility model and operational cost function of virtual power plant. This problem can be formulated as a chance constrained optimization problem and equivalently converted a deterministic convex optimization:

- Using Gaussian mixture model to characteristic the uncertainties of forecast errors;
- Transforming the risk constraints into equivalent deterministic constraints; and
- Linearizing the models of network and DERs.

Moreover, we use the convex piecewise-linear fitting algorithm to fit the stochastic PFR and cost function of VPP with the data-driven method and make it easily embedded into the system operation or market bidding model.

The results of numerical test verify the reasonableness of our model and show its higher computational efficiency than the scenario-based methods such as Monte-Carlo simulation. It can be applied in the unit commitment and obtains economic benefits.

REFERENCES

[1] D. Koraki and K. Strunz, "Wind and Solar Power Integration in Electricity Markets and Distribution Networks Through Service-Centric Virtual Power Plants," *IEEE Trans. Power Syst.*, vol. 33, no. 1, pp. 473–485, Jan. 2018, doi: 10.1109/TPWRS.2017.2710481.

[2] G. Plancke, K. De Vos, R. Belmans, and A. Delnooz, "Virtual power plants: Definition, applications and barriers to the implementation in the distribution system," in *2015 12th International Conference on the European Energy Market (EEM)*, May 2015, pp. 1–5, doi: 10.1109/EEM.2015.7216693.

[3] D. Pudjianto, C. Ramsay, and G. Strbac, "Virtual power plant and system integration of distributed energy resources," *IET Renew. Power Gener.*, vol. 1, no. 1, pp. 10–16, Mar. 2007, doi: 10.1049/iet-rpg:20060023.

[4] A. V. Jayawardena, L. G. Meegahapola, D. A. Robinson, and S. Perera, "Capability chart: A new tool for grid-tied microgrid operation," in *2014 IEEE PES T D Conference and Exposition*, Apr. 2014, pp. 1–5, doi: 10.1109/TDC.2014.6863320.

[5] A. V. Jayawardena, L. G. Meegahapola, D. A. Robinson, and S. Perera, "Microgrid capability diagram: A tool for optimal grid-tied operation,"

*Renew. Energy*, vol. 74, pp. 497–504, Feb. 2015, doi: 10.1016/j.renene.2014.08.035.
[6] S. Riaz and P. Mancarella, "On Feasibility and Flexibility Operating Regions of Virtual Power Plants and TSO/DSO Interfaces," in *2019 IEEE Milan PowerTech*, Milan, Italy, Jun. 2019, pp. 1–6, doi: 10.1109/PTC.2019.8810638.
[7] T. Sowa *et al.*, "Method for the operation planning of virtual power plants considering forecasting errors of distributed energy resources," *Electr. Eng.*, vol. 98, no. 4, pp. 347–354, Dec. 2016, doi: 10.1007/s00202-016-0419-9.
[8] Y. Yang, W. Wu, B. Wang, and M. Li, "Analytical Reformulation for Stochastic Unit Commitment Considering Wind Power Uncertainty With Gaussian Mixture Model," *IEEE Trans. Power Syst.*, vol. 35, no. 4, pp. 2769–2782, Jul. 2020, doi: 10.1109/TPWRS.2019.2960389.
[9] P. Li, D. Yu, M. Yang, and J. Wang, "Flexible Look-Ahead Dispatch Realized by Robust Optimization Considering CVaR of Wind Power," *IEEE Trans. Power Syst.*, vol. 33, no. 5, pp. 5330–5340, Sep. 2018, doi: 10.1109/TPWRS.2018.2809431.
[10] B. Zhang, R. Rajagopal, and D. Tse, "Network Risk Limiting Dispatch: Optimal Control and Price of Uncertainty," *IEEE Trans. Autom. Control*, vol. 59, no. 9, pp. 2442–2456, Sep. 2014, doi: 10.1109/TAC.2014.2325640.
[11] A. Silva *et al.*, "Assessing DER flexibility in a German distribution network for different scenarios and degrees of controllability," in *CIRED Workshop 2016*, Helsinki, Finland, 2016, pp. 1–4, doi: 10.1049/cp.2016.0801.
[12] J. Silva, J. Sumaili, R. J. Bessa, L. Seca, M. Matos, and V. Miranda, "The challenges of estimating the impact of distributed energy resources flexibility on the TSO/DSO boundary node operating points," *Comput. Oper. Res.*, vol. 96, pp. 294–304, Aug. 2018, doi: 10.1016/j.cor.2017.06.004.
[13] J. Silva *et al.*, "Estimating the Active and Reactive Power Flexibility Area at the TSO-DSO Interface," *IEEE Trans. Power Syst.*, vol. 33, no. 5, pp. 4741–4750, Sep. 2018, doi: 10.1109/TPWRS.2018.2805765.
[14] H. Chen and A. Moser, "Improved flexibility of active distribution grid by remote control of renewable energy sources," in *2017 6th International Conference on Clean Electrical Power (ICCEP)*, Santa Margherita Ligure, Italy, Jun. 2017, pp. 280–284, doi: 10.1109/ICCEP.2017.8004828.
[15] D. Mayorga Gonzalez *et al.*, "Determination of the Time-Dependent Flexibility of Active Distribution Networks to Control Their TSO-DSO Interconnection Power Flow," in *2018 Power Systems Computation Conference (PSCC)*, Dublin, Ireland, Jun. 2018, pp. 1–8, doi: 10.23919/PSCC.2018.8442865.
[16] D. A. Contreras and K. Rudion, "Improved Assessment of the Flexibility Range of Distribution Grids Using Linear Optimization," in *2018 Power Systems Computation Conference (PSCC)*, Dublin, Ireland, Jun. 2018, pp. 1–7, doi: 10.23919/PSCC.2018.8442858.
[17] D. A. Contreras and K. Rudion, "Verification of Linear Flexibility Range Assessment in Distribution Grids," in *2019 IEEE Milan PowerTech*, Jun. 2019, pp. 1–6, doi: 10.1109/PTC.2019.8810542.
[18] M. Rossi, G. Viganò, D. Moneta, M. T. Vespucci, and P. Pisciella, "Fast estimation of equivalent capability for active distribution networks," *CIRED - Open Access Proc. J.*, vol. 2017, no. 1, pp. 1763–1767, Oct. 2017, doi: 10.1049/oap-cired.2017.1273.
[19] D. A. Contreras and K. Rudion, "Impact of Grid Topology and Tap Position Changes on the Flexibility Provision from Distribution Grids," in *2019 IEEE PES Innovative Smart Grid Technologies Europe (ISGT-Europe)*, Sep. 2019, pp. 1–5, doi: 10.1109/ISGTEurope.2019.8905479.
[20] M. Cui, C. Feng, Z. Wang, and J. Zhang, "Statistical Representation of Wind Power Ramps Using a Generalized Gaussian Mixture Model," *IEEE Trans. Sustain. Energy*, vol. 9, no. 1, pp. 261–272, Jan. 2018, doi: 10.1109/TSTE.2017.2727321.
[21] A. Bernstein and E. Dall'Anese, "Linear power-flow models in multiphase distribution networks," in *2017 IEEE PES Innovative Smart Grid Technologies Conference Europe (ISGT-Europe)*, Sep. 2017, pp. 1–6, doi: 10.1109/ISGTEurope.2017.8260205.
[22] A. Bernstein, C. Wang, E. Dall'Anese, J.-Y. Le Boudec, and C. Zhao, "Load Flow in Multiphase Distribution Networks: Existence, Uniqueness, Non-Singularity and Linear Models," *IEEE Trans. Power Syst.*, vol. 33, no. 6, pp. 5832–5843, Nov. 2018, doi: 10.1109/TPWRS.2018.2823277.
[23] M. Braun, *Provision of ancillary services by distributed generators: technological and economic perspective*, vol. 10. Kassel: Kassel Univ. Press, 2009.
[24] A. Magnani and S. P. Boyd, "Convex piecewise-linear fitting," *Optim. Eng.*, vol. 10, no. 1, pp. 1–17, Mar. 2009, doi: 10.1007/s11081-008-9045-3.
[25] S. Wang and W. Wu, "Supplemental File for Stochastic Flexibility Evaluation for Virtual Power Plant by Aggregating Distributed Energy Resources." [Online]. Available: https://www.dropbox.com/s/eyepx3oxj4280wo/SupplementalFile.doc?dl=0.
[26] X. Chen, E. Dall'Anese, C. Zhao, and N. Li, "Aggregate Power Flexibility in Unbalanced Distribution Systems," *IEEE Trans. Smart Grid*, vol. 11, no. 1, pp. 258–269, Jan. 2020, doi: 10.1109/TSG.2019.2920991.
[27] H. Zhao, J. Chen, B. Wang, X. Shang, J. Zhuang, and H. Sun, "A Robust Aggregate Model for Multi-Energy Virtual Power Plant in Grid Dispatch," in *2019 IEEE Sustainable Power and Energy Conference (iSPEC)*, Beijing, China, Nov. 2019, pp. 1631–1636, doi: 10.1109/iSPEC48194.2019.8975334.
[28] CIGRE Task Force C6-04, *Benchmark systems for network integration of renewable and distributed energy resources*. Paris (21 rue d'Artois, 75008): CIGRÉ, 2014.
[29] D. Jager and A. Andreas, "NREL National Wind Technology Center (NWTC): M2 Tower; Boulder, Colorado (Data)." Not Available, 1996, doi: 10.5439/1052222.
[30] Office of Energy Efficiency & Renewable Energy (EERE), "Commercial Load Dataset," *Commercial and Residential Hourly Load Profiles for all TMY3 Locations in the United States*. https://openei.org/datasets/files/961/pub/ (accessed Jan. 16, 2020).
[31] D. Pozo, J. Contreras, and E. E. Sauma, "Unit Commitment With Ideal and Generic Energy Storage Units," *IEEE Trans. Power Syst.*, vol. 29, no. 6, pp. 2974–2984, Nov. 2014, doi: 10.1109/TPWRS.2014.2313513.